\documentclass[12pt]{iopart}

\usepackage{graphicx}
\usepackage{amssymb}
\begin{document}

\title[Extra polarization states of cosmological GWs in
alternative theories of gravity]{Extra polarization states of
cosmological gravitational waves in alternative theories of
gravity}
\author{M. E. S. Alves, O. D. Miranda and J. C. N. de Araujo}
\address{INPE-Instituto Nacional de Pesquisas Espaciais - Divis\~ao
de Astrof\'isica,
\\Av.dos Astronautas 1758, S\~ao Jos\'e dos Campos, 12227-010 SP, Brazil\\}
\ead{alvesmes@das.inpe.br, oswaldo@das.inpe.br,
jcarlos@das.inpe.br}

\begin{abstract}
Cosmological Gravitational Waves (GWs) are usually associated with
the transverse-traceless part of the metric perturbations in the
context of the theory of cosmological perturbations. These modes
are just the usual polarizations `+' and `$\times$' which appear
in the general relativity theory. However, in the majority of the
alternative theories of gravity, GWs can present more than these
two polarization states. In this context, the Newman-Penrose
formalism is particularly suitable for evaluating the number of
non-null GW modes. In the present work we intend to take into
account these extra polarization states for cosmological GWs in
alternative theories of gravity. As an application, we derive the
dynamical equations for cosmological GWs for two specific
theories, namely, a general scalar-tensor theory which presents
four polarization states and a massive bimetric theory which is in
the most general case with six polarization states for GWs.
However, the mathematical tool presented here is quite general, so
it can be used to study cosmological perturbations in all metric
theories of gravity.
\end{abstract}


\section{Introduction}

The future detection of gravitational waves (GWs) of cosmological
origin will strongly constrain the possible inflationary scenarios
which have been proposed in the last decades. Also, GWs will be
useful to distinguish between the standard inflationary model and
the alternative early Universe cosmologies, like the Pre-Big-Bang
scenario, since the predicted power spectrum of each model can
present very different features.

In the general relativity theory (GRT) the usual procedure in
order to evaluate the power spectrum of cosmological GWs is to
expand small metric perturbations around the spatially homogeneous
and isotropic Friedmann-Robsertson-Walker (FRW) metric. The next
step is to identify GWs with the transverse-traceless (TT) part of
the metric perturbations which do not couple with the
perturbations of the perfect-fluid. Thus, once it was generated,
this radiative gravitational field can freely travel through the
space and reach an observer today.

The observational effect of GWs is to generate relative tidal
accelerations between test particles. The Riemann tensor
determines these relative accelerations and it is the only locally
observable imprint of gravity. To see that, consider a freely
falling observer at any fiducial point $P$ in the region. Let the
observer set up an approximately Lorentz, normal coordinate system
$\{x^\mu\} = \{t,x^i\}$, with $P$ as origin. For a particle with
spatial coordinates $x^i$ at rest, the relative acceleration with
respect to $P$ is:
\begin{equation}\label{relative accel}
a_i = -R_{i0j0}x^j,
\end{equation}
where $R_{i0j0}$ are the so-called `electric' components of the
Riemann tensor due to waves or other external gravitational
influences. When the linearized theory is considered, the Riemann
tensor can be split in six algebraically independent components,
but for the vacuum field equations of GRT they reduce to two,
which represent the two polarization states of free GWs. These are
the `TT-modes', also called the $+$ and $\times$ polarizations.

Although these are the most studied modes in the GWs physics, when
the framework of an alternative theory of gravity is considered,
the number of non-null components of the Riemann tensor can be
greater than two and the theory presents not only the TT-modes,
but also other polarization states can appear. This is a direct
consequence of the new field equations which can generate other
radiative modes. Thus, for a generic theory of gravity, GWs can
present up to six polarization states corresponding to the six
independent components of $R_{i0j0}$.

Therefore, we should state that in order to work out the
cosmological metric perturbations in an alternative theory of
gravity, the first step is to find the number of independent
polarizations of GWs in such a theory, i.e., the number of
non-null components of the Riemann tensor. This can be done by a
very elegant method which consists in the evaluation of the
non-null Newman-Penrose (NP) quantities \cite{Eardley1973,Newman}
of a given theory. These quantities are the irreducible parts of
the Riemann tensor written in a complex tetrad basis which make
them very useful to evaluate the polarizations of GWs in an
unambiguous way.

Thus, the present paper intends to consider the general formalism
of cosmological perturbations in the context of alternative
theories of gravity, focusing on the dynamical equations of the GW
modes. The theory of cosmological perturbations in GRT has been
largely studied in the literature. Some classical examples are the
works by Lifshiftz \cite{Lifshiftz1946}, Bardeen
\cite{Bardeen1980}, Peebles \cite{Peebles1993}, and Mukhanov,
Feldman and Brandenberger \cite{Mukhanov1992}. For a recent review
on cosmological dynamics see, e.g.,\cite{Durrer2004,Malik2009}. In
the usual approach, the TT-modes of GWs are consistently described
by the superadiabatic, or parametric, amplification mechanism in
the GRT \cite{Grishchuk1974}. Furthermore, it was shown by Barrow
and de Garcia Maia \cite{Barrow1993_2,Maia1994} that the same
mechanism applies for modified theories of gravity as
scalar-tensor theories and the so-called $f(R)$ theories. Although
they have analyzed only the evolution of the two TT-modes, it is
known since the work by Eardley et al. \cite{Eardley1973} that
scalar-tensor theories present in addition at least one scalar GW
polarization (more general scalar-tensor theories present two
scalar GW polarizations) as a consequence of the additional degree
of freedom included by the scalar field. The relic scalar GW
production which arises from this kind of theories was recently
discussed by Capozziello et al. \cite{Capozziello2007}, and an
upper limit was obtained from the amplitude of scalar
perturbations in the Wilkinson Microwave Anisotropy Probe (WMAP)
data. But in their analysis, Capozziello et al. have considered
only the vacuum field equations, this is a limitation of their
results since a coupling between scalar GWs and the scalar
perturbations of the cosmological perfect fluid are expected as
will be clear in our derivations.

In the case of $f(R)$ theories, using the NP formalism, it was
shown that a particular class of functions $f(R)$ presents two
scalar GW modes in addition to the $+$ and $\times$ modes, thus
totalizing four independent polarizations of GWs
\cite{Alves2009B}. However, when the Palatini approach is used in
the derivation of the field equations, the theory reveals only the
usual two TT polarizations. It was also found that the scalar
longitudinal mode which appears in $f(R)$ theories is a massive
mode which is potentially  detectable by the future space GW
interferometer LISA \cite{Capozziello2008}. Again by considering
only the vacuum equations, the production of the relic GWs of this
particular mode was also considered and constraints using the WMAP
data was established \cite{Corda2008}.

Moreover, the study of extra polarization states of cosmological
GWs in the context of alternative theories of gravity can reveal
new interesting features of these theories which do not appear in
GRT. A remarkable example is the presence of vector longitudinal
polarization modes of GWs in some theories. These modes give rise
to a non usual Sachs-Wolf effect which leaves a vector signature
on the CMB polarization \cite{Bessada2008}. Otherwise, vector
perturbations in GRT decay too fast and it would not leave any
signature on CMB polarization.

Therefore, it is clear that the future detection of GWs, and the
corresponding determination of the number of polarization modes,
are powerful tools to test the underlying gravity theory. Thus,
the goal of the present paper is to furnish a general formalism to
find the evolution equations of all the possible polarization
modes which could appear in a generic theory of gravity. Once the
number of independent polarization modes are found and the
corresponding evolutionary equations could be obtained, one is in
a position to obtain the power spectrum of each mode, finding the
CMB signatures and constraining the additional modes. In order to
show the application of the formalism, and in order to find some
new features of the current studied theories, we have chosen to
obtain the dynamical equations for GWs in the context of two
particular theories, namely, a general scalar-tensor theory and a
bimetric massive theory of gravity.

First proposed by Brans and Dicke \cite{Brans1961} in the aim of
making the theory of gravity compatible with the Mach's principle,
the scalar-tensor theories are of a great interest since, as
pointed out by several authors, a coupling between a scalar field
and gravity seems to be a generic outcome of the low-energy limit
of string theories (see, e.g., \cite{Casas1991}). Another interest
in the scalar-tensor models is that the $f(R)$ theories can be
written as the Einstein equations plus a scalar field, and thus we
could in principle extend the same formalism applied for the
scalar-tensor theories to the $f(R)$ field equations. The bimetric
massive theory we consider was proposed by Visser \cite{vis1998}
in the aim to obtain general covariant field equations with
massive gravitons. His method was based on the introduction of a
non-dynamical metric $(g_0)_{\mu\nu}$ besides the physical metric
$g_{\mu\nu}$. The resulting equations appear as a small
modification of the Einstein field equations for which the massive
gravitons and the metric $(g_0)_{\mu\nu}$ are present only in an
additional energy-momentum tensor. Furthermore, our past studies
have shown that the Visser's theory is a potential explanation for
the current acceleration of the expansion of the Universe
\cite{Alves2006,Alves2009}.

In deriving the equations for GWs in the two theories we will
first review how to obtain the number of independent polarization
modes for any theory following the Eardley et al. approach
\cite{Eardley1973}. In the case of the scalar-tensor models the
theory present four polarization states in the more general case.
Otherwise, the Visser's theory is a simple example of how a weak
modification of gravity can produce six polarization modes. The
subsequent analysis show that all the polarization modes, apart
from the usual $+$ and $\times$ polarizations, are dynamically
``coupled'' to the perturbations of the cosmological perfect
fluid. We argue that this kind of coupling and the existence of
additional polarization states could furnish distinguishable
signatures of alternative theories in the power spectrum of the
relic GWs.

The paper is organized as follows: in the section \ref{sec 0} we
present an overview of the NP formalism starting from the
definition of the NP quantities which define the six possible
polarization states for GWs. Then we find the non-vanishing
parameters for the GRT, scalar-tensor theories and for the
Visser's model. In the section \ref{sec 2}, considering a generic
theory, we find general expressions for the perturbed Einstein
tensor and for the energy-momentum tensor in the generalized
harmonic coordinates. In the section \ref{sec 3} we introduce a
decomposition scheme which depends on the number of non-vanishing
polarization modes of GWs which could appear in the various
alternative theories. In the sections \ref{sec 4} and \ref{sec 5}
we apply the formalism of the preceding sections for two
particular theories, the scalar-tensor theory and the Visse's
bimetric model. Finally, we present our conclusions and
discussions in the section \ref{sec 6}.

Throughout the paper we use units such that $c=1$ unless otherwise
mentioned.

\section{An overview of the Newman-Penrose formalism}\label{sec 0}

\subsection{Tetrads components of tensors and null tetrads}

At every point of the space one can introduce systems of four
linearly independent vectors $e_\mu^{(a)}$, which are known as
tetrads. The index in parenthesis is the tetrad index which
numbers the vectors from one to four. We can define the matrix:
\begin{equation}\label{eq2}
g^{(a)(b)}=e_\mu^{(a)}e_\nu^{(b)}g^{\mu\nu},
\end{equation}
which is an arbitrary symmetric matrix with negative-definite
determinant. Its inverse $g_{(c)(a)}$, which is defined by:
\begin{equation}
g_{(c)(a)}g^{(a)(b)} = \delta^{(b)}_{(c)},
\end{equation}
can be used to lower the tetrad indices:
\begin{equation}\label{eq3}
e_{(a)\mu}=g_{(a)(b)}e^{(b)}_{(\mu)},
\end{equation}
and to solve (\ref{eq2}) for $g_{\mu\nu}$:
\begin{equation}
g_{\mu\nu} = g_{(a)(b)}e^{(a)}_\mu e^{(b)}_\nu.
\end{equation}

One can write any arbitrary vector or tensor as a linear
combination of the four tetrad vectors:
\begin{equation}
{T^{\mu\nu \cdot \cdot \cdot}}_{\gamma\sigma \cdot \cdot \cdot} = {T^{(a)(b)\cdot \cdot \cdot}}_{(c)(d)\cdot \cdot \cdot}
e^\mu_{(a)} e^\nu_{(b)} e_\gamma^{(c)} e_\sigma^{(d)}\cdot \cdot \cdot,
\end{equation}
where the quantities ${T^{(a)(b)\cdot \cdot \cdot}}_{(c)(d)\cdot
\cdot \cdot}$ are the tetrad components of the tensor. They are
calculated according to:
\begin{equation}\label{tetrad components}
{T^{(a)(b)\cdot \cdot \cdot}}_{(c)(d)\cdot \cdot \cdot} = {T^{\mu\nu \cdot \cdot \cdot}}_{\gamma\sigma \cdot \cdot \cdot}
e_\mu^{(a)} e_\nu^{(b)} e^\gamma_{(c)} e^\sigma_{(d)}\cdot \cdot \cdot,
\end{equation}
which is consistent with (\ref{eq2}) and (\ref{eq3}). Tetrad
indices are raised and lowered with $g^{(a)(b)}$ and $g_{(a)(b)}$
respectively.

The advantages offered in many cases by the use of the tetrad
components become clear when one examines their transformation
properties and when one introduces tetrads which are appropriate
to the particular problem being investigated. From the equation
(\ref{tetrad components}) one can see that the tetrad components
behave like scalars under coordinate transformations, i.e., the
tetrad indices of tensors do not change under a coordinate
transformation. Therefore, we have a good way of investigating the
algebraic properties of tensors in a coordinate-independent
fashion by the choice of the tetrads.

A possible choice is to identify the tetrad vectors with the base
vectors of a Cartesian coordinate system in the local Minkowski
system of the point concerned:
\begin{equation}
g_{(a)(b)} = e^\mu_{(a)} e^\nu_{(b)}g_{\mu\nu} = \eta_{(a)(b)} = {\rm{diag}}(-1,1,1,1).
\end{equation}

The four tetrad vectors, which we shall call $(\textbf{e}_t,
\textbf{e}_x, \textbf{e}_y, \textbf{e}_z)$, form an orthonormal
system of one timelike and three spacelike vectors.

Another special case is the use of null vectors as tetrad vectors.
A particular tetrad, known as Newman-Penrose tetrad \cite{Newman}
can be constructed from the orthonormal system introduced above,
two of the four vectors $\textbf{k}$ and $\textbf{l}$ are real
null vectors:
\begin{equation}
\textbf{k} = \frac{1}{\sqrt{2}}( \textbf{e}_t + \textbf{e}_z ),~~\textbf{l} = \frac{1}{\sqrt{2}}( \textbf{e}_t - \textbf{e}_z ),
\end{equation}
and the other two null vectors $\textbf{m}$ and
$\overline{\textbf{m}}$ are compex conjugates of each other:
\begin{equation}
\textbf{m} = \frac{1}{\sqrt{2}}( \textbf{e}_x + i \textbf{e}_y ),~~\overline{\textbf{m}} = \frac{1}{\sqrt{2}}( \textbf{e}_x - i\textbf{e}_y ).
\end{equation}

It is easy to verify that the tetrad vectors obey the relations:
\begin{equation}
-\textbf{k} \cdot \textbf{l} = \textbf{m} \cdot \overline{\textbf{m}} = 1
\end{equation}
\begin{equation}
\textbf{k} \cdot \textbf{m} = \textbf{k} \cdot \overline{\textbf{m}} =
\textbf{l} \cdot \textbf{m} = \textbf{l} \cdot \overline{\textbf{m}} = 0,
\end{equation}
and from (\ref{eq2}) and (\ref{eq3}) we obtain:
 \begin{equation}
 g_{(a)(b)} = \left(
   \begin{array}{cccc}
     0  &  1  &  0  &  0  \\
     1  &  0  &  0  &  0  \\
     0  &  0  &  0  &  -1 \\
     0  &  0  &  -1 &  0
   \end{array}
              \right).
 \end{equation}

A null tetrad basis is especially suitable for discussing null or
nearly null waves.

\subsection{Tetrad components of the Riemann tensor}

The Riemann tensor $R_{\lambda\mu\nu\kappa}$ can be split in the
irreducible parts: the Weyl tensor, the traceless Ricci tensor and
the curvature scalar (see, e.g., \cite{Weinberg1972}), whose
tetrad components can be named respectively as $\Psi$, $\Phi$ and
$\Lambda$ following the notation of \cite{Newman}. In general, in
a four dimensional space we have ten $\Psi$'s, nine $\Phi$'s and
one $\Lambda$ which are all algebraically independent. However,
when we restrict ourselves to nearly plane waves, we find that the
differential and algebraic properties of $R_{\lambda\mu\nu\kappa}$
reduce the number of independent components to six
\cite{Eardley1973}. Thus, following Eardley et al.
\cite{Eardley1973} we shall choose the set
$\{\Psi_2,\Psi_3,\Psi_4,\Phi_{22}\}$ to describe, in a given null
frame, the six independent components of a wave in a generic
metric theory. These NP quantities are related to the following
components of the Riemann tensor in the null tetrad basis
described earlier:
\begin{equation}
\Psi_2 = -\frac{1}{6} R_{lklk},
\end{equation}
\begin{equation}
\Psi_3 = -\frac{1}{2} R_{lkl\overline{m}},
\end{equation}
\begin{equation}
\Psi_4 = - R_{l\overline{m}l\overline{m}},
\end{equation}
\begin{equation}
\Phi_{22} = - R_{lml\overline{m}}.
\end{equation}

\begin{figure}[!ht]
\centering
\includegraphics[width=90mm]{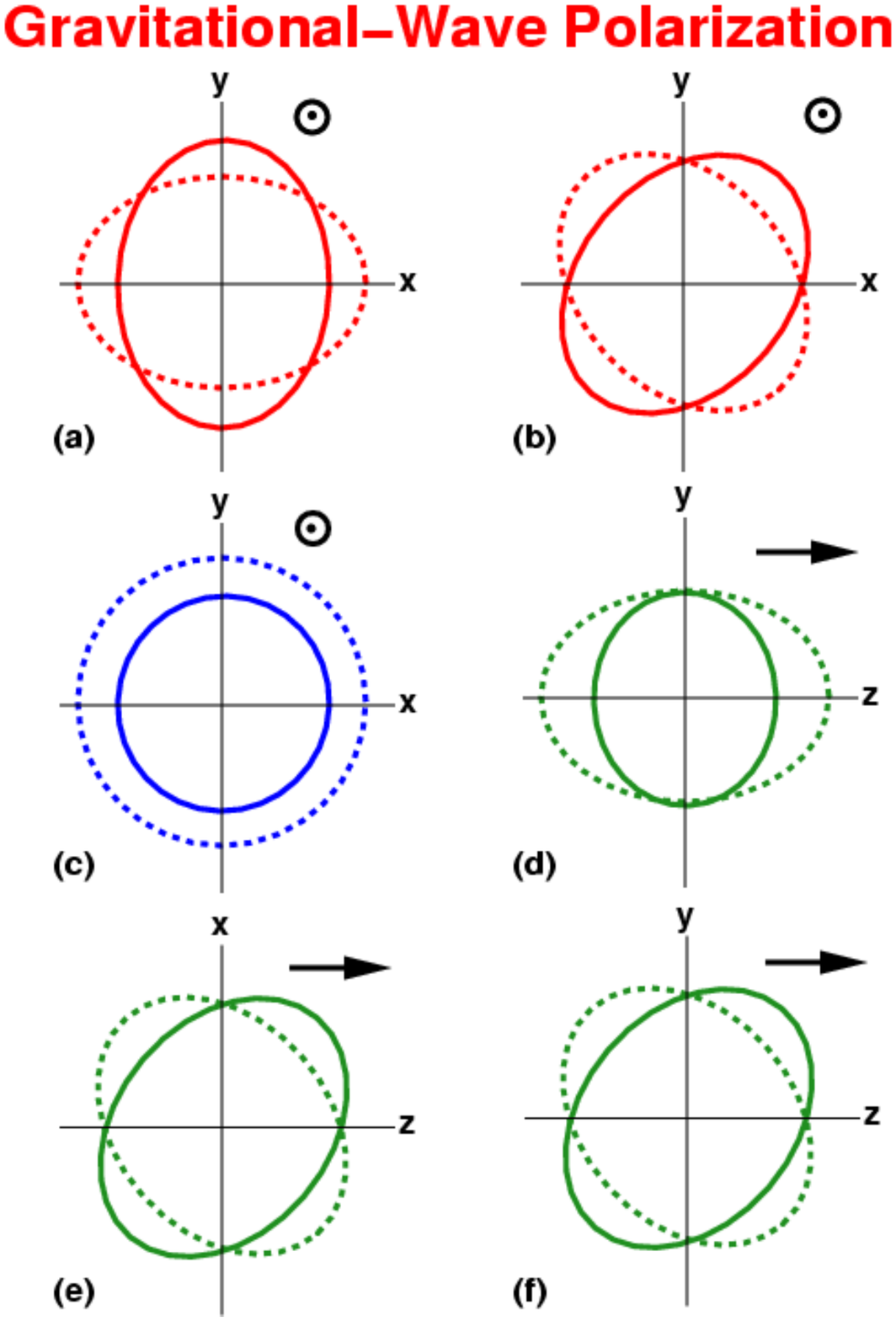}
\caption{The six polarization modes of weak, plane,
null GW permitted in any metric theory of gravity.
Also shown is the displacement that each mode induces on a sphere of test particles. The wave
propagates out of the plane in $(a)$, $(b)$ and $(c)$, and it propagates in the plane
in $(d)$, $(e)$ and $(f)$. The displacement induced on the sphere of test particles
corresponds to the following Newman–Penrose quantities: Re$\Psi_4(a)$, Im$\Psi_4(b)$,
$\Phi_{22}(c)$, $\Psi_2(d)$, Re$\Psi_3(e)$, Im$\Psi_3(f)$. See ref. \cite{Will2006}.}
\label{fig1}
\end{figure}

Note that, $\Psi_3$ and $\Psi_4$ are complex, thus each one
represents two independent polarizations. One polarization for the
real part and one for the imaginary part, thus totalizing six
components. Three are transverse to the direction of propagation,
with two representing quadrupolar deformations and one
representing a monopolar ``breathing'' deformation. Three modes
are longitudinal, with one an axially symmetric stretching mode in
the propagation direction, and one quadrupolar mode in each of the
two orthogonal planes containing the propagation direction. The
Fig. \ref{fig1}, which was taken from \cite{Will2006}, shows the
displacements induced on a ring of freely falling test particles
by each one of these modes. GRT predicts only the first two
transverse quadrupolar modes (a) and (b).

Other useful expressions are the following relations for the Ricci
tensor:
\begin{equation}\label{ricci 1}
R_{lk} = R_{lklk},
\end{equation}
\begin{equation}
R_{ll} = 2R_{lml\overline{m}},
\end{equation}
\begin{equation}
R_{lm} = R_{lklm},
\end{equation}
\begin{equation}\label{ricci 4}
R_{l\overline{m}} = R_{lkl\overline{m}},
\end{equation}
and for the curvature scalar:
\begin{equation}\label{curv scalar}
R = -2R_{lk} = -2R_{lklk}.
\end{equation}

The overall relative accelerations in a sphere of test particles
is described by the relation (\ref{relative accel}) and can be
expressed in terms of the symmetric ``driving-force matrix''
$\textbf{S}$ whose components are not but the electric components
of the Riemann tensor $S_{ij} = R_{i0j0}$, where the latin indices
represent spatial coordinates. These components can be written as
a combination of the NP quantities in the following way:
\begin{equation}
  \textbf{S} = \left(
    \begin{array}{ccc}
      -\frac{1}{2}({\rm{Re}}\Psi_4 + \Phi_{22})  &  \frac{1}{2}{\rm{Im}}\Psi_4  &  -2{\rm{Re}}\Psi_3  \\
      \frac{1}{2}{\rm{Im}}\Psi_4                 &  \frac{1}{2}({\rm{Re}}\Psi_4 - \Phi_{22})  & 2{\rm{Im}}\Psi_3 \\
      -2{\rm{Re}}\Psi_3                          &  2{\rm{Im}}\Psi_3            &  -6\Psi_2
    \end{array}
                \right).
\end{equation}
Since each NP amplitude is linearly independent, we can expand the
components $S_{ij}$ as the sum:
\begin{equation}
S_{ij} = \sum_{r = 1}^6 p_{(r)} E^{(r)}_{ij},
\end{equation}
where we have renamed the NP quantities as follows
\cite{Eardley1973}:
\begin{equation}
p_{(1)} \equiv \Psi_2,
\end{equation}
\begin{equation}
p_{(2)} \equiv {\rm{Re}} \Psi_3,
\end{equation}
\begin{equation}
p_{(3)} \equiv {\rm{Im}} \Psi_3,
\end{equation}
\begin{equation}
p_{(4)} \equiv {\rm{Re}} \Psi_4,
\end{equation}
\begin{equation}
p_{(5)} \equiv {\rm{Im}} \Psi_4,
\end{equation}
\begin{equation}
p_{(6)} \equiv \Phi_{22},
\end{equation}
and $E^{(r)}_{ij}$ are the components of the ``basis polarization
matrices'' which are given by:.
\begin{equation}\label{basis polarization matrices}
   \begin{array}{cc}
      \mathbf{E}_{(1)} = -6\left(
        \begin{array}{ccc}
          0 & 0 & 0 \\
          0 & 0 & 0 \\
          0 & 0 & 1
        \end{array}
      \right) &

      \mathbf{E}_{(2)} = -2\left(
        \begin{array}{ccc}
          0 & 0 & 1 \\
          0 & 0 & 0 \\
          1 & 0 & 0
        \end{array}
      \right) \\
      \\

      \mathbf{E}_{(3)} = 2\left(
        \begin{array}{ccc}
          0 & 0 & 0 \\
          0 & 0 & 1 \\
          0 & 1 & 0
        \end{array}
      \right) &

      \mathbf{E}_{(4)} = -\frac{1}{2}\left(
        \begin{array}{ccc}
          1 & 0 & 0 \\
          0 & -1 & 0 \\
          0 & 0 & 0
        \end{array}
      \right) \\
      \\

      \mathbf{E}_{(5)} = \frac{1}{2}\left(
        \begin{array}{ccc}
          0 & 1 & 0 \\
          1 & 0 & 0 \\
          0 & 0 & 0
        \end{array}
      \right) &

      \mathbf{E}_{(6)} = -\frac{1}{2}\left(
        \begin{array}{ccc}
          1 & 0 & 0 \\
          0 & 1 & 0 \\
          0 & 0 & 0
        \end{array}
      \right).

   \end{array}
 \end{equation}

Finally, analyzing the behavior of the set
$\{\Psi_2,\Psi_3,\Psi_4,\Phi_{22}\}$ under rotations, we see that
they have the respective helicity values $s=\{0, \pm 1, \pm 2, 0
\}$. These are all the possible helicity values for GWs in a
general theory of gravitation.

\subsection{Determining the polarizations of GWs for some specific
theories}

The procedure of evaluation of the number of independent
polarizations involves examining the far-field, linearized, vacuum
field equations of a theory, and then finding the non-null NP
amplitudes. For GW sources sufficiently far from the observer we
hope that the external solutions of GWs approach that of vacuum
and linearized regime, in this way the method becomes an
unambiguous tool for the determination of the propagating physical
modes of GWs. In what follows we are going to evaluate the NP
components of the Riemann tensor for three theories, namely, the
GRT, a general class of scalar-tensor theories and a bimetric
massive theory.

In the case of the GRT, the field equations can be obtained from
the Einsten-Hilbert action:
\begin{equation}
I = \frac{1}{16\pi G}\int \sqrt{-g}R d^4x  + I_M,
\end{equation}
where $I_M$ is the action which describes the matter fields. From
the Hamilton principle $\delta I = 0$ and the Einstein equations
read:
\begin{equation}\label{einsten equations}
G_{\mu\nu} = -8\pi G T_{\mu\nu}.
\end{equation}

For the vacuum $T_{\mu\nu} = 0$ and (\ref{einsten equations})
reduces to:
\begin{equation}
R_{\mu\nu} = 0.
\end{equation}

Therefore, from the relations for the Ricci tensor and for the
curvature scalar in the NP tetrad basis (\ref{ricci 1})-(\ref{curv
scalar}) we can show that:
\begin{equation}
R_{lklk} = R_{lkl\overline{m}} = R_{lml\overline{m}} = 0,
\end{equation}
or equivalently:
\begin{equation}
\Psi_2 = \Psi_3 = \Phi_{22} = 0,
\end{equation}
and since we have no further constraints on the components of the
Riemann tensor we conclude that:
\begin{equation}
\Psi_4 \neq 0.
\end{equation}
Hence, as expected, a GW in the GRT presents two polarization
states with helicity $s = \pm 2$.
\\
\\
The scalar-tensor theories were first proposed by Brans and Dicke
\cite{Brans1961} in the aim of making the theory of gravity
compatible with the Mach's principle. These theories are of great
interest since a coupling between a scalar field and gravity seems
to be a generic outcome of the low-energy limit of string theories
(see, e.g., \cite{Casas1991}). Here, we will consider a class of
theories characterized by the general action discussed by Wagoner
\cite{Wagoner1970} and Nordtvedt \cite{Nordtvedt1970}, which we
write as found in Maggiore and Nicolis \cite{Maggiore2000}:
\begin{equation}\label{scalar-tens lagrang}
I = \frac{1}{16\pi}\int \sqrt{-g}d^4 x\left[ -\varphi R
+ \varphi^{-1}\omega(\varphi)\nabla_\alpha\varphi\nabla^\alpha\varphi - U(\varphi) \right]+ I_M,
\end{equation}
where $\varphi$ is a scalar field, $\omega(\varphi)$ is the
coupling parameter and $U(\varphi)$ can be interpreted as a
potential associated with $\varphi$. The Brans-Dicke action can be
obtained for the special case when $\omega$ is constant and
$U(\varphi)=0$. The field equations obtained by minimizing
(\ref{scalar-tens lagrang}) with respect to the metric and with
respect to the scalar field are:
\begin{equation}\label{einstein phi eq}
G_{\mu\nu} = -\frac{8\pi }{\varphi} T_{\mu\nu} - \frac{\omega}{\varphi^2}
              \left(\varphi_{;\mu}\varphi_{;\nu} - \frac{1}{2} g_{\mu\nu}
              \varphi_{;\alpha}\varphi^{;\alpha} \right)
              -\frac{1}{\varphi}(\varphi_{;\mu\nu}-g_{\mu\nu}\Box \varphi) - Ug_{\mu\nu},
\end{equation}
\begin{equation}\label{phi equation}
\left[3 + 2\omega(\varphi)\right]\Box \varphi -\varphi \frac{dU}{d\varphi}= 8\pi  T
- \frac{d\omega}{d\varphi}\varphi_{;\alpha}\varphi^{;\alpha},
\end{equation}
\begin{equation}\label{cons energ st}
{{T_\mu}^\nu}_{;\nu} = 0.
\end{equation}

Note that the last condition, which express the conservation of
the energy-momentum tensor, must be independently imposed since it
is not a direct consequence of the field equations as in the case
of the GRT. The vacuum field equations now read:
\begin{equation}\label{ricci again}
R_{\mu\nu} = -\frac{\omega(\varphi)}{\varphi^2}\varphi_{;\mu}\varphi_{;\nu}
- \frac{1}{\varphi}\left(\varphi_{;\mu\nu} + \frac{1}{2}g_{\mu\nu}\Box \varphi\right) + g_{\mu\nu}U(\varphi),
\end{equation}
and
\begin{equation}\label{scalar ricci st}
R = -\frac{\omega}{\varphi^2}\varphi_{;\alpha}\varphi^{;\alpha} - 3\frac{\Box{\varphi}}{\varphi}+4U.
\end{equation}

Let us first evaluate the polarizations for the case of the
Brans-Dicke theory, for which the equation (\ref{phi equation})
becomes simply:
\begin{equation}
\Box \varphi = 0,
\end{equation}
with the solution:
\begin{equation}\label{solução BD}
\varphi = \varphi_0 + \varphi_1 e^{iq_\alpha x^\alpha},
\end{equation}
where $\varphi_0$ is a constant obtained from the cosmological
boundary conditions, $\varphi_1$ is a small amplitude in such a
way we can work only to first order in $\varphi_1$, and
$q_{\alpha}$ is the wave vector which is null for this particular
case. It follows that the curvature scalar is null:
\begin{equation}
R = 0 \Rightarrow R_{lklk} = 0,
\end{equation}
and the Ricci tensor takes the form:
\begin{equation}
R_{\mu\nu} = -\frac{\varphi_1}{\varphi_0} e^{iq_\alpha x^\alpha}q_{\mu}q_{\nu}.
\end{equation}

Hence if the wave is propagating in the $+z$ direction, the only
non-null components of the Ricci tensor are $R_{zz}$, $R_{zt}$ and
$R_{tt}$, which means that $R_{ll}$ is the only non-null component
in the tetrad basis, leading us to conclude that:
\begin{equation}
R_{lkl\overline{m}} = 0 ~ {\rm{and}} ~ R_{lml\overline{m}} \neq 0,
\end{equation}
and therefore:
\begin{equation}
\Psi_2 = \Psi_3 = 0, ~ \Psi_4 ~ {\rm{and}} ~ \Phi_{22} \neq 0.
\end{equation}

That is, GWs in the Brans-Dicke theory presents the two $s= \pm 2$
polarizations of GRT plus a perpendicular breathing polarization
mode with helicity $s = 0$.

Otherwise, if the potential $U(\varphi)$ is not null we have a
more general scalar-tensor theory. Now, $\Box \varphi$ is not null
in general, but obeys the following relation in the linearized
regime \cite{Maggiore2000}:
\begin{equation}\label{linearized gen ST}
\Box \varphi -m_0^2 \varphi = 0,
\end{equation}
where:
\begin{equation}
m_0^2 \equiv \frac{\varphi_0\left(dU/d\varphi\right)_{\varphi=\varphi_0}}{3 + 2\omega(\varphi_0)},
\end{equation}
and $\varphi_0$ is the value of $\varphi$ for which the potential
minimum is evaluated. The solution of (\ref{linearized gen ST}) is
given by (\ref{solução BD}), but now the wave vector satisfies:
\begin{equation}
q_\alpha q^\alpha = - m_0^2,
\end{equation}
thus the curvature scalar does not vanish but gives:
\begin{equation}
R = -3m^2_0\frac{\varphi_1}{\varphi_0}e^{iq_\alpha x^\alpha},
\end{equation}
for first order in $\varphi_1$, and the equation for the Ricci
tensor is:
\begin{equation}
R_{\mu\nu} = -\frac{\varphi_1}{\varphi_0}e^{iq_\alpha x^\alpha}
\left[ q_\mu q_\nu + \frac{1}{2}\eta_{\mu\nu}m_0^2\right].
\end{equation}

Therefore, from the evaluation of the tetrad components of the
Ricci tensor and from the preceding relations for the Riemann
tensor, we conclude that for a general scalar-tensor theory:
\begin{equation}
\Psi_3 = 0;~\Psi_2,~\Psi_4~{\rm{and}}~\Phi_{22} \neq 0.
\end{equation}

Now, we have a longitudinal scalar polarization mode besides the
three modes which appear in the Brans-Dicke theory. Note also that
the appearance of the longitudinal mode is related to the presence
of a massive scalar field in the theory.
\\
\\
The next theory we are going to analyze is an alternative theory
of gravity which takes into account massive gravitons by the
action \cite{vis1998}:
\begin{equation}\label{massive action}
I =  \int d^4x \left[\frac{\sqrt{-g}R}{16\pi G} + m^2 {\cal{L}}_{mass}(g,g_0)\right] + I_M,
\end{equation}
where $m$ is the graviton mass in natural units and the Lagrangian
${\cal{L}}_{mass}$ is a function of the physical metric $g$ and of
a non-dynamical prior defined metric $g_0$. The massive Lagrangian
proposed by Visser is:
\begin{eqnarray}\label{massive lagrangian}
{\cal{L}}_{mass}(g,g_0) =& \frac{1}{2}\sqrt{-g_0}\bigg\{ ( g_0^{-1})^{\mu\nu} ( g-g_0)_{\mu\sigma}( g_0^{-1})^{\sigma\rho} \nonumber \\
  &\times ( g-g_0)_{\rho\nu}-\frac{1}{2}\left[( g_0^{-1})^{\mu\nu}( g-g_0)_{\mu\nu}\right]^2\bigg\}.
\end{eqnarray}

By varying the action (\ref{massive action}) with respect to the
physical metric we obtain the field equations:
 \begin{equation}\label{field-equations visser}
   G^{\mu\nu} -\frac{1}{2}m^2 M^{\mu\nu} = -8\pi G  T^{\mu\nu},
 \end{equation}
where the massive tensor $M^{\mu\nu}$ reads:
 \begin{equation}\label{massive tensor}
   M^{\mu\nu} =  (g_0^{-1})^{\mu\sigma}\bigg[ (g-g_0)_{\sigma\rho}-\frac{1}{2}(g_0)_{\sigma\rho}(g_0^{-1})^{\alpha\beta}
   (g-g_0)_{\alpha\beta} \bigg](g_0^{-1})^{\rho\nu}.
 \end{equation}

Among the existing bimetric theories of gravity (Rosen's bimetric
theory \cite{Rosen1973}, for example), the most common criterion
for the choice of the non-dynamical background metric is to impose
the Riemann-flat condition
${R^\lambda}_{\mu\nu\kappa}(\textbf{g}_0) = 0$
\cite{Will2006,Will1993}. Thus, the simpler choice is a flat
metric which recovers the Minkowski metric by going to cartesian
coordinates \cite{vis1998,Alves2006}. By introducing this
background metric, Visser has constructed the Lagrangian
(\ref{massive lagrangian}) aiming a general covariant description
of massive gravitons in such a way that the model circumvents the
van Dam-Veltman-Zakharov discontinuity
\cite{vDVZ1970,Zakharov1970}, a inconsistency which plagues other
massive terms.

Furthermore, the Visser's model could be an alternative
explanation to the current acceleration of the expansion of the
Universe as indicated by the recession of distant Supernova
\cite{Alves2009}. In this sense, the massive tensor $M_{\mu\nu}$
mimics the dark energy effects in large scales while the theory
passes all the local tests of gravity. Hence, it is a matter of
interest to distinguish between the Visser's theory and the
Einstein gravity by dynamical tests such as GW observations.

From (\ref{field-equations visser}) we see that the vacuum field
equations with massive gravitons read:
\begin{equation}\label{visser vacuo}
R^{\mu\nu} = m^2 \left(M^{\mu\nu} - \frac{1}{2} \eta^{\mu\nu}M\right),
\end{equation}
and for the weak field approximation the right-hand-side of
(\ref{visser vacuo}) is simply $m^2 h_{\mu\nu}$ where $h_{\mu\nu}$
is a metric perturbation around the flat background metric
$g_{\mu\nu} = (g_0)_{\mu\nu} + h_{\mu\nu}$ with $|h_{\mu\nu}|\ll
1$. Therefore, since there are no further restrictions on
$R_{\mu\nu}$, we conclude that all its components are non-null and
we find that:
\begin{equation}
R_{lklk},~R_{lml\overline{m}},~R_{lklm}~{\rm{and}}~R_{lkl\overline{m}} \neq 0,
\end{equation}
and then:
\begin{equation}
\Psi_2,~\Psi_3,~\Psi_4~{\rm{and}}~\Phi_{22} \neq 0.
\end{equation}

Thus, GWs in the Visser's theory present all the six possible
polarization states showed in the Fig. \ref{fig1}. This result was
first obtained by de Paula et al. \cite{paula2004} in a different
but equivalent approach. Note that in the limit $m\rightarrow 0$,
the polarization modes $\Psi_2$, $\Psi_3$ and $\Phi_{22}$ vanish
and we recover the GRT with the only non-null mode $\Psi_4$.

\section{Cosmological Perturbations without decomposition}\label{sec
2}

Let us consider a general theory of gravity for which the field
equations can be written in the form:
\begin{equation}\label{general field eq.}
G_{\mu\nu} + F_{\mu\nu} = - 8\pi G T_{\mu\nu},
\end{equation}
where besides the Einstein tensor $G_{\mu\nu}$ and the
energy-momentum tensor for the matter fields $T_{\mu\nu}$, we have
a general tensor
$F_{\mu\nu}=F_{\mu\nu}(g^{\alpha\beta},\gamma^{\alpha\beta},\varpi^\alpha,\varphi,\ldots)$
which could be a function of the physical metric $g_{\mu\nu}$, of
some prior defined metric $\gamma_{\mu\nu}$, of vector fields
$\varpi_\mu$, of scalar fields $\varphi$ and of derivatives of
these quantities. A prior defined metric is in general considered
in bimetric theories of gravity, for which besides the dynamical
metric $g_{\mu\nu}$, a kind of ``absolute'' geometry is specified
through $\gamma_{\mu\nu}$. One of the most known theory of this
kind is the Rosen's theory \cite{Rosen1973} for which the second
metric takes into account the effects of inertial forces. Another
example is the bimetric massive theory considered by Visser which
was introduced in the last section and we will analyze in more
detail in the section \ref{sec 5}.

Supposing that there is a cosmological solution of such a theory,
we can work out the perturbations on a cosmological background
metric $g_{\mu\nu}$. We adopt the metric $g_{\mu\nu}$ as the
Robertson-Walker metric written in Cartesian coordinates with zero
spatial curvature $k=0$. With these considerations the line
element reads:
\begin{equation}\label{RW metric}
ds^2 = a^2(\eta)\eta_{\mu\nu}dx^\mu dx^\nu,
\end{equation}
where the scale factor $a(\eta)$ is a function of the conformal
time $x^0=\eta$. The cosmic time $t$ is related to the conformal
time by the relation $a(\eta)d\eta=dt$ and the Minkowski metric is
$\eta_{\mu\nu} = {\rm{diag}}(-1,+1,+1,+1)$.

Perturbing the metric (\ref{RW metric}) in the general form
$\delta g_{\mu\nu} = a^2(\eta)h_{\mu\nu}(x^\alpha)$ with
$|h_{\mu\nu}|\ll 1$, we now have:
\begin{equation}\label{pert metric}
ds^2 = a^2(\eta)(\eta_{\mu\nu} + h_{\mu\nu})dx^\mu dx^\nu.
\end{equation}

In this form, the indices of $h_{\mu\nu}$ are raised and lowered
by the metric $\eta_{\mu\nu}$. With the line element (\ref{pert
metric}) in the equations (\ref{general field eq.}) we can obtain
the perturbed field equations:
\begin{equation}\label{pert. field equations}
\delta G_{\mu\nu} + \delta F_{\mu\nu} = - 8\pi \left[ G \delta T_{\mu\nu} + \delta G T_{\mu\nu}\right],
\end{equation}
where, in order to take into account the theories with varying
Newtonian ``constant'', we have included the perturbation $\delta
G$.

Hereafter we will use the generalized harmonic coordinates
discussed by Bicak and Katz \cite{Bicak2005}. For this coordinate
system, we have the condition:
\begin{equation}\label{harm coord.}
g^{\mu\nu}\delta \Gamma^\lambda_{\mu\nu} = 0,
\end{equation}
where $\delta \Gamma^\lambda_{\mu\nu}$ is the metric connection
perturbation. It is easy to show that, the condition (\ref{harm
coord.}) is equivalent to:
\begin{equation}\label{gauge}
\nabla_\nu \delta \bar{g}^{\mu\nu} = 0,
\end{equation}
where $\delta \bar{g}_{\mu\nu} = \delta g_{\mu\nu} - g_{\mu\nu}
\delta g/2$ with $\delta g = g^{\alpha\beta}\delta
g_{\alpha\beta}$. One can show that $\delta \bar{g}_{\mu\nu}=a^2
\bar{h}_{\mu\nu}$ where the $\bar{h}_{\mu\nu}$ is the
trace-reverse perturbation:
\begin{equation}
\bar{h}_{\mu\nu} = h_{\mu\nu} - \frac{1}{2}\eta_{\mu\nu} h,
\end{equation}
with $h = \eta^{\mu\nu}h_{\mu\nu}$ and $\bar{h}=-h$.

With the metric perturbations given by (\ref{pert metric}) and
with the condition (\ref{gauge}) we can evaluate the components of
the perturbed Einstein tensor $\delta G_\mu^\nu$ for first order
in $h_{\mu\nu}$. A straightforward calculation leads to:
 \begin{eqnarray}\label{pert einst 1}
   \delta G_0^0 = \frac{1}{2a^2}\Big[& - \bar{h}_0^{0\prime\prime} + \nabla^2\bar{h}_0^0  -2 {\cal{H}}\bar{h}_0^{0\prime}
                  + 3(3{\cal{H}}^2-{\cal{H}}^{\prime})\bar{h}_0^0  \nonumber \\
                                     &- ({\cal{H}}^2 - {\cal{H}}^{\prime} )\bar{h}_i^i + 4a{\cal{H}} \partial_i \bar{h}_0^i  \Big],
 \end{eqnarray}
 \begin{eqnarray}\label{pert einst 2}
   \delta G_0^i = \frac{1}{2a^2}\Big[& - \bar{h}_0^{i\prime\prime} + \nabla^2\bar{h}_0^i - 4{\cal{H}}\bar{h}_0^{i\prime}
                  +({\cal{H}}^2 - {\cal{H}}^\prime)\bar{h}_0^i \nonumber \\
                                     & + 2a^{-1}{\cal{H}}\eta^{ij}(\partial_j \bar{h}_0^0 - \partial_k \bar{h}_j^k) \Big],
 \end{eqnarray}
 \begin{eqnarray}\label{pert einst 3}
   \delta G_i^j = \frac{1}{2a^2}\Big[& - \bar{h}_i^{j\prime\prime} + \nabla^2\bar{h}_i^j - 2{\cal{H}}\bar{h}_i^{j\prime}
                  + ({\cal{H}}^2 + {\cal{H}}^\prime)\bar{h}_k^k \delta_i^j \nonumber \\
                                     & - ({\cal{H}}^2 - {\cal{H}}^\prime)\bar{h}_0^0\delta_i^j
                                       - 4a{\cal{H}} \eta^{jk}\partial_{(k}\bar{h}_{i)0} \Big].
 \end{eqnarray}

The prime in the above expressions denotes derivatives with
respect to the conformal time $\eta$, and we have defined the
Hubble parameter for the conformal time as ${\cal{H}} \equiv
a^{\prime}/a $.

Now, it is necessary to evaluate the perturbations of the
energy-momentum tensor for a perfect fluid:
\begin{equation}
T_{\mu\nu} = (\rho+P)U_\mu U_\nu+Pg_{\mu\nu},
\end{equation}
where $\rho$ and $P$ are the energy density and the pressure, and
$U_{\nu} = - U^{\nu} = (1,0,0,0)$ is the fluid four-velocity.
Considering first order perturbations in each one of these
quantities the perturbed energy-momentum tensor reads:
\begin{equation}
T_{\mu\nu} \rightarrow \tilde{T}_{\mu\nu} = T_{\mu\nu} + \delta T_{\mu\nu},
\end{equation}
where:
\begin{eqnarray}
   \delta {T_\mu}^\nu = & (\rho + P)g^{\lambda \nu}(U_\mu \delta U_\lambda
   + \delta U_\mu U_\lambda) + (\delta \rho + \delta P) U_\mu U^\nu
   + \delta P {\delta_\mu}^\nu \nonumber \\
                        &- (\rho + P) \delta g^{\lambda \nu} U_\mu U_\lambda.
\end{eqnarray}

And considering the perturbed metric as defined earlier we find
the components of the perturbed quantity $\delta {T_\mu}^\nu$:
\begin{equation}
\delta T_0^0 = -\delta \rho,~~~\delta T_0^i = (\rho + P)(V^i + a^{-2}\bar{h}_0^i),~~~\delta T_i^j = \delta P \delta_i^j,
\end{equation}
where, for completeness, we have considered $V^i = \delta U^i \ll
1$ as the spatial part of the perturbation of the fluid
four-velocity as it appears in the references
\cite{Mukhanov1992,Bicak2007}:
\begin{equation}
\tilde{U}^\nu = U^\nu + \delta U^\nu = (1-\frac{1}{2}\delta g_{00},~V^i).
\end{equation}

It is easy to see that the four-velocity is approximately the unit
timelike vector since we assume $V^i \ll 1$, and terms
proportional to $V^2$ and $Vh$ can thus be neglected.

Then, from the calculation of the first order perturbation of the
tensor $\delta F_{\mu\nu}$ and using the perturbed Einstein tensor
and the perturbed energy-momentum tensor given above, one can
write the perturbed field equations for a generic theory (see eq.
\ref{pert. field equations}).

As we will see in the following sections, the above equations are
written in a suitable way to consider extra-polarization states of
the GWs in any alternative theory of gravity with the field
equations (\ref{general field eq.}). They are also useful to find
solutions without any decomposition as it was presented in the
interesting work by Bicak et al. \cite{Bicak2007}. Note that,
despite of the difference in notation, a similar set of equations
can be found in that reference.

\section{Metric perturbations with extra polarization states of
GWs}\label{sec 3}

Instead of considering the minimal decomposition which appears,
for example, in the reference \cite{Mukhanov1992}, let us write
the components of $\delta g_{\mu\nu}$ in a more general form which
was first presented by Bessada and Miranda \cite{Bessada2008}:
  \begin{equation}
    \delta \bar{g}_{\mu\nu} = a^2(\eta)
    \left(
      \begin{array}{cc}
         \phi                  &    S_i + \partial_i B \\
         S_i + \partial_i B    &    \bar{h}_{ij}
      \end{array}
    \right),
  \end{equation}
where $\phi$ and $B$ are scalar perturbations and $S_i$ are the
components of a divergenceless vector perturbation $\partial_i S^i
= 0$. In our approach, the quantities $\phi$, $B$ and $S_i$ are
dynamical quantities which appear in any theory. On the other
hand, the way the perturbation $\bar{h}_{ij}$ must be decomposed
depends on the number of independent GW polarization modes, and
hence, it is theory dependent. Thus, the first step is to evaluate
the number of non-null NP quantities and then we can decompose
$\bar{h}_{ij}$ identifying the GW amplitudes in the expansion. In
order to introduce the method, let us first consider the most
general case for which all the NP quantities are non-null. For
this particular case, $\bar{h}_{ij}$ can be expanded in terms of
six components which represent the six metric amplitudes of GWs:
\begin{equation}\label{expansion}
\bar{h}_{ij}(\textbf{x},\eta) = \sum_{r=1}^6 \epsilon_{ij}^{(r)}\bar{h}_{(r)}(\textbf{x},\eta),
\end{equation}
where $\epsilon_{\mu\nu}^{(r)}$ are the polarization tensors.

In what follows, without loss of generality, we can consider the
wave vector of the GW oriented in the $+z$ direction. Thus, we can
construct the six $\epsilon_{\mu\nu}^{(r)}$ by combinations of the
three ortonormal vectors:
\begin{equation}\label{base vectors}
   \begin{array}{lcl}
     \ell^i & = & \left(1,0,0\right)\\
     m^i & = & \left(0,1,0\right)\\
     n^i & = & \left(0,0,1\right),
   \end{array}
 \end{equation}
in the following way:
 \begin{equation}\label{polar 1}
   \epsilon^{ij}_{\tiny{\left(1\right)}} = n^i n^i~,
 \end{equation}
 \begin{equation}
   \epsilon^{ij}_{(2)} = \ell^i n^j + \ell^j n^i~,
 \end{equation}
 \begin{equation}
   \epsilon^{ij}_{(3)} = m^i n^j + m^j n^i~,
 \end{equation}
 \begin{equation}
   \epsilon^{ij}_{(4)} = \ell^i \ell^j - m^i m^j~,
 \end{equation}
  \begin{equation}
   \epsilon^{ij}_{(5)} = \ell^i m^j + \ell^j m^i~,
 \end{equation}
 \begin{equation}\label{polar 6}
   \epsilon^{ij}_{(6)} = \ell^i \ell^j + m^i m^j~.
 \end{equation}

One can verify that the tensors (\ref{polar 1}) - (\ref{polar 6})
are linearly independent and form an ortogonal basis. Writing in a
matricial form we have:
\begin{equation}\label{polarization matrices}
   \begin{array}{cc}
      \left[\epsilon^{ij}_{(1)}\right] = \left(
        \begin{array}{ccc}
          0 & 0 & 0 \\
          0 & 0 & 0 \\
          0 & 0 & 1
        \end{array}
      \right) &

      \left[\epsilon^{ij}_{(2)}\right] = \left(
        \begin{array}{ccc}
          0 & 0 & 1 \\
          0 & 0 & 0 \\
          1 & 0 & 0
        \end{array}
      \right) \\
      \\

      \left[\epsilon^{ij}_{(3)}\right] = \left(
        \begin{array}{ccc}
          0 & 0 & 0 \\
          0 & 0 & 1 \\
          0 & 1 & 0
        \end{array}
      \right) &

      \left[\epsilon^{ij}_{(4)}\right] = \left(
        \begin{array}{ccc}
          1 & 0 & 0 \\
          0 & -1 & 0 \\
          0 & 0 & 0
        \end{array}
      \right) \\
      \\

      \left[\epsilon^{ij}_{(5)}\right] = \left(
        \begin{array}{ccc}
          0 & 1 & 0 \\
          1 & 0 & 0 \\
          0 & 0 & 0
        \end{array}
      \right) &

      \left[\epsilon^{ij}_{(6)}\right] = \left(
        \begin{array}{ccc}
          1 & 0 & 0 \\
          0 & 1 & 0 \\
          0 & 0 & 0
        \end{array}
      \right).

   \end{array}
 \end{equation}

We can see that, apart from a constant, these are just the basis
polarization matrices (\ref{basis polarization matrices}) which we
have described earlier. Now, the expansion (\ref{expansion}) can
be written in a more intuitive form:
\begin{equation}
\bar{h}_{ij} = \pi_{ij} + \tau_{ij} + \chi_{ij} + \psi_{ij},
\end{equation}
where we have defined four new tensors following their helicity
values $s$:
\begin{equation}\label{definition pert.}
\begin{array}{lcl}
\pi_{ij}  \equiv \bar{h}_{ij}^{(1)} ~~~&\rightarrow & s=0\\
\tau_{ij} \equiv \bar{h}_{ij}^{(2)} + \bar{h}_{ij}^{(3)} ~~~ &\rightarrow & s=\pm 1\\
\chi_{ij} \equiv \bar{h}_{ij}^{(4)} + \bar{h}_{ij}^{(5)} ~~~ &\rightarrow & s=\pm 2\\
\psi_{ij} \equiv \bar{h}_{ij}^{(6)} ~~~&\rightarrow & s=0
\end{array}
\end{equation}

And from the very definition of the quantities (\ref{definition
pert.}), we have the following properties:
\begin{equation}
\chi_i^i = \tau_i^i = 0,~~~~\partial_j\chi_i^j = \partial_j\psi_i^j = 0.
\end{equation}

Their associated helicity values can be evaluated from the
behaviour of the perturbations under rotations. Notice also that
we have chosen to define separately the two tensors for the $s =
0$ modes of GWs since one of them is a longitudinal mode
($\pi_{ij}$) and the other is transversal to the direction of
propagation ($\psi_{ij}$). Furthermore, we have seen in the
section \ref{sec 0} that these two modes can appear for some
theories (in a general scalar-tensor theory, for example), but
other theories have a structure such that only one scalar mode
appears (this is the case of the Brans-Dicke theory). This is a
particular feature of the scalar modes, since it is not possible
that a certain theory presents only one of the two tensor modes
and not the other. In fact, there is no theory for which the $+$
and $\times$ polarizations appear separated for vacuum GWs as it
was evidenced by the construction of the NP amplitude $\Psi_4$
(see section \ref{sec 0}). The same happens with the two vector
modes which generate the same NP quantity $\Psi_3$.

It is instructive to write the NP quantities in terms of these
metric perturbations considering an observer today located at the
local Minkowskian space-time. Thus, considering the Cartesian
tetrad basis we have:
\begin{equation}
\Psi_2 = -\frac{1}{12}(\pi_{zz,00} - 2 \pi_{zz,zz} + \phi_{,zz}) + \frac{1}{24}(\bar{h}_{,00}-\bar{h}_{,zz}),
\end{equation}
\begin{equation}
\Psi_3 = \frac{1}{4\sqrt{2}} [(\tau_{xz,zz} - \tau_{xz,00}) + i (\tau_{yz,00} - \tau_{yz,zz}) ],
\end{equation}
\begin{equation}
\Psi_4 = \frac{1}{4} [\chi_{yy,00} - \chi_{xx,00} + 2i \chi_{xy,00}],
\end{equation}
\begin{equation}
\Phi_{22} = -\frac{1}{4}(\psi_{xx,00} + \psi_{yy,00} - \bar{h}_{,00}),
\end{equation}
where:
\begin{equation}
\bar{h} = - \phi + \psi^i_i + \pi^i_i.
\end{equation}

Now, since the polarization modes of GWs are linearly independent,
we can summarize the procedure to calculate the perturbed field
equations for a given general theory of the form (\ref{general
field eq.}) in the following way:
\\

\begin{itemize}
\item \emph{\textbf{Scalar metric perturbations}}
\end{itemize}

The general scalar metric perturbations are
  \begin{equation}
    \delta {{\bar{g}}^{(s)}}_{\mu\nu} = a^2(\eta)
    \left(
      \begin{array}{cc}
         \phi            &     \partial_i B \\
         \partial_i B    &     h_{ij}^{(s)}
      \end{array}
    \right).
  \end{equation}

If the theory has $\Psi_2\neq 0$ and $\Phi_{22}\neq 0$, it means
that an observer today can measure the two scalar GW modes ($s =
0$). Hence, in order to describe the evolution of these modes we
write the term $h_{ij}^{(s)}$ in the form:
\begin{equation}
h_{ij}^{(s)} = \pi_{ij} + \psi_{ij},
\end{equation}
with the constraint $\partial_j \psi^j_i = 0$ which guarantees the
transversal propagation of the tensor $\psi^j_i$.

If the theory has $\Psi_{2} = 0$ and $\Phi_{22}\neq 0$, no
longitudinal scalar GW amplitude can be measured. Then $\pi_{ij}$
is suppressed, and in order to do not change the number of degrees
of freedom of the metric perturbations, a scalar longitudinal
component, say $D$, must be added. This new scalar component
represents a dynamical perturbation of the cosmic gravitational
potential rather than a radiative GW field. Consequently, an
observer today measures only one scalar GW mode whose evolution
can be described by $\psi_{ij}$, and $h_{ij}^{(s)}$ takes the
form:
\begin{equation}
h_{ij}^{(s)} = \partial_i\partial_j D + \psi_{ij},
\end{equation}
where we have introduced $D$ using the spatial partial derivatives
in order to represent the longitudinal behaviour of this quantity.
This is the same procedure of introducing longitudinal scalars in
the minimal decomposition \cite{Mukhanov1992}. For those theories
which can be found in this case (e.g., the Brans-Dicke theory),
the only non-null scalar degree of freedom (in the local
Minkowskian frame) is $\psi_{ij}$. In this sense, it is always
possible to make $D = 0$ by gauge transformations in the reference
frame of the local far field observer, although in general $D$
does not cancel out when considering the whole cosmological
evolution of the perturbations.

On the other hand, if the theory has $\Psi_2=\Phi_{22}=0$, the
perturbations must be decomposed in the usual way, since there are
no scalar GWs which could be observed today, that is:
\begin{equation}
h_{ij}^{(s)} = E\delta_{ij} + \partial_i\partial_jD,
\end{equation}
where we have added the new scalar quantity $E$ to describe the
evolution of the perturbations of the gravitational potential. The
most important theory which can be found in this case is the
Einstein theory for which we have only the two ``pure tensor''
degrees of freedom for vacuum GWs. In Minkowski coordinates, it is
always possible to cancel out all the scalar components by gauge
transformations, but in the presence of the cosmic fluid they are
important to describe the evolution of the perturbations.
\\

\begin{itemize}
\item \emph{\textbf{Vector metric perturbations}}
\end{itemize}

The general vector metric perturbations for any theory are:
  \begin{equation}
    \delta {{\bar{g}}^{(v)}}_{\mu\nu} = a^2(\eta)
    \left(
      \begin{array}{cc}
         0      &         S_i     \\
         S_i    &     h_{ij}^{(v)}
      \end{array}
    \right),
  \end{equation}
where the vector $S_i$ and $h_{ij}^{(v)}$ satisfies:
\begin{equation}\label{vector condition}
\partial_i S^i = h_i^{i(v)} =  0
\end{equation}

If the NP quantity $\Psi_3$ is non-null (the Visser's theory is an
example), an observer today can measure the two vector modes ($s =
\pm 1$) of GWs, and the metric perturbations $h_{ij}^{(v)}$ can be
identified to the corresponding GW amplitude:
\begin{equation}
h_{ij}^{(v)} = \tau_{ij},
\end{equation}
with $\tau_i^i = 0$.

On the other hand, if $\Psi_3 = 0$, there are no vector GWs today
and we have the usual representation in terms of the vector
quantity $Q_i$:
\begin{equation}
h_{ij}^{(v)} = \partial_i Q_j + \partial_j Q_i,
\end{equation}
where, from (\ref{vector condition}) $Q_i$ is divergenceless
$\partial_i Q^i = 0$ and again, it is easy to verify that the
number of the degrees of freedom of the metric perturbations does
not change. The quantity $Q_i$ is a vector perturbation of the
gravitational potential for which the dynamical equations gives,
in general, a decaying mode in the context of the GRT (see, e.g.
\cite{Bardeen1980,Mukhanov1992}). As in the case of scalar
perturbations we have discussed earlier, it is always possible to
cancel $Q_i$ by gauge transformations in the local Minkowski frame
when $\Psi_3 = 0$.
\\

\begin{itemize}
\item \emph{\textbf{Tensor metric perturbations}}
\end{itemize}

Finally, the tensor metric perturbations are constructed using a
symmetric tensor $\chi_{ij}$ which satisfies the constraints:
\begin{equation}
\chi_i^i = \partial_j\chi_i^j = 0.
\end{equation}

Thus, the tensor component which corresponds to GWs with $s=\pm 2$
is written in the usual way in a theory-independent form:
  \begin{equation}
    \delta {{\bar{g}}^{(t)}}_{\mu\nu} = a^2(\eta)
    \left(
      \begin{array}{cc}
         0      &         0      \\
         0      &     \chi_{ij}
      \end{array}
    \right).
  \end{equation}

Counting the number of independent components we have used to
construct $\delta g_{\mu\nu}$, and the number of constraints, we
can see that we have four functions for scalar perturbations, four
functions for vector perturbations, and two functions for tensor
perturbations. Thus, as expected, we have ten independent
components of $\delta g_{\mu\nu}$.

In the context of the Einstein theory we have already found that
$\Psi_4$ is the only non-null NP quantity, thus it is direct to
verify that the metric perturbations have the usual minimal
decomposition:
 \begin{equation}\label{GRT perturbations}
    \delta \bar{g}_{\mu\nu} = a^2(\eta)
    \left(
      \begin{array}{cc}
         \phi                  &                      S_i + \partial_i B \\
         S_i + \partial_i B    &    E\delta_{ij} + \partial_i\partial_jD + \partial_i Q_j + \partial_j Q_i + \chi_{ij}
      \end{array}
    \right),
  \end{equation}
and GWs are described only by the quantity $\chi_{ij}$ whose
evolution equation can be derived from $\delta G_{\mu\nu}^{(t)} =
-8\pi G T_{\mu\nu}^{(t)}$ to obtain the very known result:
  \begin{equation}\label{GRT tensor pert}
    \chi_i^{j\prime\prime} + 2{\cal{H}}\chi_i^{j\prime} - \nabla^2 \chi_i^j = 0.
  \end{equation}

In the following sections we will exemplify the above
decomposition scheme for the two other theories which we have
treated in the section \ref{sec 0}, namely, a general
scalar-tensor theory and the Visser's bimetric theory with massive
gravitons. For each theory, after identifying the form of the
metric perturbations we will find the dynamical equations for GWs
in the coordinate system defined by (\ref{harm coord.}).

\section{GWs in scalar-tensor theories}\label{sec 4}

In this section we will consider perturbations of the general
class of scalar-tensor theories introduced in the section \ref{sec
0}. With a glance at the field equations (\ref{einstein phi eq})
we identify it with the general form (\ref{general field eq.})
where $G(\varphi) = \phi^{-1}$ and the generic function
$F_{\mu\nu}$ takes the form:
\begin{equation}
F_{\mu\nu} = \frac{\omega(\varphi)}{\varphi^2}
              \left(\varphi_{;\mu}\varphi_{;\nu} - \frac{1}{2} g_{\mu\nu}
              \varphi_{;\alpha}\varphi^{;\alpha} \right)
              +\frac{1}{\varphi}(\varphi_{;\mu\nu}-g_{\mu\nu}\Box \varphi)
              + g_{\mu\nu}U(\varphi).
\end{equation}

Thus, disturbing the Newtonian ``constant'':
\begin{equation}
\delta G = \frac{dG}{d\varphi}\delta\varphi = -\frac{\delta \varphi}{\varphi^2},
\end{equation}
and evaluating the components of the perturbation $\delta
F_{\mu\nu}$, we can find the perturbed field equations (\ref{pert.
field equations}) which for this case reads:
\begin{equation}
\delta G_{\mu\nu} = -\frac{8\pi}{\varphi}\left(\delta T_{\mu\nu} - \frac{\delta \varphi}{\varphi}T_{\mu\nu}\right)
+\delta F_{\mu\nu}.
\end{equation}

As we have already shown in the section \ref{sec 0}, the
evaluation of the NP parameters for the most general scalar-tensor
theory lead us to conclude that an observer today would measure:
\begin{equation}
\Psi_2 \neq 0,~~\Psi_3 = 0,~~\Psi_4 \neq 0~~{\rm{and}}~~\Phi_{22} \neq 0,
\end{equation}
remembering that $\Psi_2 = 0$ if the potential associated to the
scalar field is null $U(\varphi) = 0$. Thus, following the
procedure of the last section we can write the perturbations for
the general case which reads:
 \begin{equation}\label{Scalar-tensor perturbations}
    \delta \bar{g}_{\mu\nu} = a^2(\eta)
    \left(
      \begin{array}{cc}
         \phi                  &                      S_i + \partial_i B \\
         S_i + \partial_i B    &    \pi_{ij} + \chi_{ij} + \psi_{ij} + \partial_i Q_j + \partial_j Q_i
      \end{array}
    \right).
  \end{equation}

Now, we can introduce the perturbations given above in the
perturbed Einstein tensor and in the perturbed energy-momentum
tensor calculated without decomposition in the section \ref{sec
2}, and we can also evaluate the perturbed quantity $\delta
F_{\mu\nu}$. Thus, after a straightforward calculation we find the
dynamical equations which describe scalar and tensorial GWs in the
context of the scalar-tensor theories:
\\
\\
\emph{Scalar}
  \begin{eqnarray}\label{ST scalar 1}
    \phi^{\prime\prime} + 2 {\cal{H}}\phi^{\prime} -
    (9{\cal{H}}^2-3{\cal{H}}^{\prime})\phi - \nabla^2 \phi  - \Big[\frac{\omega}{2}\Big(\frac{\varphi^{\prime}}{\varphi}\Big)^2
    + \frac{\varphi^{\prime\prime}}{\varphi}\Big](\phi +\xi^i_i)  \nonumber \\
    -\frac{1}{2}[({\cal{H}}^{\prime}-{\cal{H}}^2)]\xi_i^i
    - 2a{\cal{H}}\nabla^2 B   =
    16\pi a^2\varphi^{-1} \Big( - \delta \rho + \frac{\delta \varphi}{\varphi}\rho\Big)    - 2 \Delta_1,
  \end{eqnarray}
  \begin{eqnarray}\label{ST scalar 2}
    (\partial^iB)^{\prime\prime} + 4{\cal{H}}(\partial^iB)^{\prime} - \nabla^2(\partial^iB)
    + ({\cal{H}}^{\prime}-{\cal{H}}^2) \partial^iB  + \Big[\frac{\varphi^{\prime\prime}}{\varphi}
    + \omega \Big(\frac{\varphi^\prime}{\varphi}\Big)^2 \Big] \partial^i B  \nonumber \\
    - 2{\cal{H}}a^{-1}(2\partial^i \phi + \partial_k
    \pi^{ik}) = 16\pi G a^2(\rho + P)(V_{\parallel}^i + \partial^i B) + 2\Delta_2^i,
  \end{eqnarray}
  \begin{eqnarray}\label{ST scalar 3}
    \xi_i^{j\prime\prime} + 2 {\cal{H}}\xi_i^{j\prime}- \nabla^2 \xi_i^j + ({\cal{H}}^\prime + {\cal{H}}^2)\xi_k^k \delta_i^j
   - 4{\cal{H}}a(\partial_i\partial^jB) \nonumber \\
    + ({\cal{H}}^\prime - {\cal{H}}^2) \phi\delta_i^j = \frac{16\pi a^2}{\varphi}\Big[\delta P\delta_i^j - \frac{\delta \varphi}{\varphi}
    P\delta_i^j\Big] +   2 {\Delta_{3i}}^{ j} \nonumber \\
    +\frac{\varphi^\prime}{\varphi} \Big[  {\cal{H}}\Big(\overline{\xi}_i^j + \frac{1}{2}\delta_i^j \phi\Big)
    +\frac{1}{4}\omega \frac{\varphi^\prime}{\varphi}(\phi+\xi_k^k)\delta_i^j  \Big]
  \end{eqnarray}
\\
\\
\emph{Tensor}
  \begin{equation}\label{ST tensor}
    \chi_i^{j\prime\prime} + \left(2{\cal{H}} + \frac{\varphi^{\prime}}{\varphi} \right) \chi_i^{j\prime}
    + 2 {\cal{H}}\frac{\varphi^{\prime}}{\varphi}\chi_i^j - \nabla^2 \chi_i^j = 0.
  \end{equation}

In the above equations we defined $\xi_i^j = \pi_i^j + \psi_i^j$,
and $V^i_\parallel$ is the component of $V^i$ which is parallel to
the direction of propagation. The perturbed quantities $\Delta_1$,
$\Delta_2^i$ and ${\Delta_{3i}}^j$ take into account the
perturbation of the scalar field $\varphi$ and its derivatives.
They are defined as follows:
  \begin{equation}
    \Delta_1 = \omega \frac{\varphi^\prime}{\varphi}\left(\frac{\delta \varphi}{\varphi} \right)^\prime
    + \frac{1}{2}\left(\frac{\varphi^\prime}{\varphi}\right)^2 \delta \omega
    + 3 {\cal{H}}\frac{\varphi^\prime}{\varphi}\frac{\delta \varphi}{\varphi}
    + a^2\left[\frac{\delta({\varphi_{;00}})}{\varphi} +\frac{\delta (\Box \varphi)}{\varphi}\right],
  \end{equation}

  \begin{equation}
    {\Delta_2}^i = \omega \frac{\varphi^\prime}{\varphi}\partial^i\left(\frac{\delta \varphi}{\varphi}\right)
    + \delta^{ij}\frac{\delta({\varphi_{;0j}})}{\varphi},
  \end{equation}

  \begin{eqnarray}
    {\Delta_{3i}}^j &=& \omega \frac{\varphi^\prime}{\varphi}\Big(\frac{\delta\varphi}{\varphi}\Big)^\prime\delta_i^j
    + \frac{1}{2}\Big(\frac{\varphi^\prime}{\varphi}\Big)^2\delta\omega \delta_i^j
    +\Big(a^2\frac{\Box \varphi}{\varphi} + {\cal{H}} {\frac{\varphi^\prime}{\varphi}}\Big)\frac{\delta \varphi}{\varphi}\delta_i^j \nonumber \\
    &+&\frac{1}{\varphi}\Big[\delta(\varphi_{;il})\delta^{lj} - a^2 \delta(\Box \varphi)\delta_i^j\Big].
  \end{eqnarray}

The equation for tensor perturbations in scalar-tensor theories
(\ref{ST tensor}) was studied in the reference
\cite{Barrow1993_2}. It represents the evolution of free GWs with
helicity $s = \pm 2$ in such theories. These modes do not couple
with the perturbations of the perfect fluid and are generated
quantum-mechanically due to vacuum perturbations in the very early
Universe. They are amplified by the expansion of the Universe due
the process known as superadiabatic amplification
\cite{Grishchuk1974}. In this case, the scalar field $\varphi$
contributes for the cosmological potential which generates the
amplification.

Regarding scalar perturbations, the driven equations for the NP
modes $\Psi_2$ and $\Phi_{22}$ are given by the equations (\ref{ST
scalar 1}), (\ref{ST scalar 2}) and (\ref{ST scalar 3}). These set
of equations, considerably more complicated than the tensorial
case, show some new physical features when compared with the usual
metric decomposition. The most important difference is that these
equations represent the evolution of radiative fields coupled to
the evolution of the generalized Newtonian gravitational
potential. The radiative fields are represented by the quantities
$\pi_{ij}$ and $\psi_{ij}$ while the generalized Newtonian
potential is described by the dynamical perturbations $\phi$ and
$B$ or some particular combination of them. Thus, in some sense,
we can say that if the scalar-tensor theory is the ``correct''
theory we would have a new kind of cosmological GW background. On
the contrary to the tensorial case, this new background is coupled
to the matter perturbations of the perfect fluid. This is expected
since the usual scalar perturbations are also coupled to the fluid
dynamics. But, in the usual sense, GWs would be coupled to the
matter only if a component of anisotropic stress would be present.

A direct consequence of such a coupling is that the whole
evolution of the density perturbations of the cosmic fluid would
depend not only on the evolution of $\phi$ and $B$ but also on the
evolution of the amplitudes of the scalar GWs $\pi_{ij}$ and
$\psi_{ij}$. Thus, from (\ref{ST scalar 1}), for example, we are
lead to conclude that:
\begin{equation}
\frac{\delta \rho}{\rho} = \frac{\delta \rho}{\rho}(\phi,B,\xi^i_i,\delta \varphi),
\end{equation}
where we have also included the dependence on the perturbations of
the scalar field $\delta \varphi$. Notice that, in fact, the
dependence of $\delta \rho/\rho$ with the GWs amplitudes appears
as the dependence on the trace of the overall contribution of the
scalar GW amplitudes $\xi^i_i = \pi^i_i + \psi^i_i$. Therefore, a
complete understanding of the evolution of the GWs with helicity
$s = 0$ requires the knowledge of the evolution of the scalar
perturbations and, similarly, these GW modes affect the evolution
of the density perturbations.

Another issue of particular interest is how the presence of the
scalar GWs would affect the angular pattern of the CMB. Since the
GW amplitudes $\pi_{ij}$ and $\psi_{ij}$ now enter the geodesic
equation for photons, it is expected that they leave a signature
on the CMB due the so-called Sachs-Wolfe effect, which can be
understood as the shift of photon frequency along the line of
sight. The small fluctuations in the CMB may be conveniently
described by perturbations of the temperature parameter $T$ in the
Planck distribution $f$. The computation of the contribution of
the scalar GWs to the angular temperature inhomogeneities $\delta
T/T$ of the CMB is out of the scope of the present paper and a
rigorous treatment will appear elsewhere.

\section{GW modes in a bimetric theory of gravity}\label{sec 5}

Now, let us turn our attention to the massive bimetric theory
first considered by Visser \cite{vis1998} which we have introduced
in the section \ref{sec 0}. Comparing the field equations
(\ref{field-equations visser}) with the generic form (\ref{general
field eq.}) we identify $F^{\mu\nu} = -m^2 M^{\mu\nu}$.

Our explicit calculations of the section \ref{sec 0}, and the
previous result by de Paula et al. \cite{paula2004}, have led us
to the conclusion that all the NP quantities are non-null for the
Visser's model:
\begin{equation}
\Psi_2 \neq 0,~~\Psi_3 \neq 0,~~\Psi_4 \neq 0~~{\rm{and}}~~\Phi_{22} \neq 0.
\end{equation}

Thus, following the procedure of the section \ref{sec 3}, the
perturbations now should be written in the form:
 \begin{equation}\label{Visser perturbations}
    \delta \bar{g}_{\mu\nu} = a^2(\eta)
    \left(
      \begin{array}{cc}
         \phi                  &                      S_i + \partial_i B \\
         S_i + \partial_i B    &         \pi_{ij} + \tau_{ij} + \chi_{ij} + \psi_{ij}
      \end{array}
    \right).
  \end{equation}

With (\ref{Visser perturbations}) in the perturbed field equations
(\ref{pert. field equations}), calculating  the components of
$\delta F_{\mu\nu}$ and with the help of the perturbed components
of the Einstein tensor and of the energy-momentum tensor
calculated in the section (\ref{sec 2}), we obtain the perturbed
field equations for Visser's theory for each group of
perturbations:
\\
\\
\emph{Scalar}
  \begin{eqnarray}{\label{Vis scalar 1}}
    \phi^{\prime\prime} + 2 {\cal{H}}\phi^{\prime} - \nabla^2 \phi -
    (9{\cal{H}}^2-3{\cal{H}}^{\prime}-m^2a^2)\phi   \nonumber \\
    -[({\cal{H}}^{\prime}-{\cal{H}}^2)-m^2a^2(a^2-1)]\xi_i^i
    -4a{\cal{H}}\nabla^2 B = - 16\pi G a^2\delta \rho,
  \end{eqnarray}
  \begin{eqnarray}\label{Vis scalar 2}
    (\partial^iB)^{\prime\prime} + 4{\cal{H}}(\partial^iB)^{\prime} - \nabla^2(\partial^iB)
    + \frac{1}{2}[2({\cal{H}}^{\prime}-{\cal{H}}^2) + m^2 a^4(3-a^2)]\partial^iB \nonumber \\
    - 2{\cal{H}}a^{-1}(2\partial^i \phi + \partial_k
    \pi^{ik}) = 16\pi G a^2(\rho + P)(V_{\parallel}^i + \partial^i B),
  \end{eqnarray}
  \begin{eqnarray}\label{Vis scalar 3}
    \xi_i^{j\prime\prime} + 2 {\cal{H}}\xi_i^{j\prime}- \nabla^2 \xi_i^j - ({\cal{H}}^\prime + {\cal{H}}^2)\xi \delta_i^j
    +\frac{1}{2}m^2 a^4(a^2 +1)\xi_i^j  \nonumber \\
    + 4{\cal{H}}a(\partial_i\partial^jB)
    - \frac{1}{2}[2({\cal{H}}^\prime - {\cal{H}}^2) - m^2 a^4(a^2 - 1)] \phi\delta_i^j = 16\pi G a^2 \delta P \delta_i^j.
  \end{eqnarray}
\\
\\
\emph{Vector}
  \begin{eqnarray}\label{Vis vector 1}
    S^{i\prime\prime} + 4 {\cal{H}} S^{i\prime} - \nabla^2 S^i + \frac{1}{2}[2({\cal{H}}^\prime - {\cal{H}}^2) - m^2a^4(a^2 - 3)]S^i \nonumber \\
    +2{\cal{H}}a^{-1}\partial_k \tau^{ik} = 16\pi G a^2(\rho + P)(V^i_{\perp} + S^i),
  \end{eqnarray}
  \begin{eqnarray}\label{Vis vector 2}
    \tau_i^{j\prime\prime} + 2{\cal{H}}\tau_i^{j\prime} - \nabla^2 \tau_i^j +
    \frac{1}{2}m^2 a^4(a^2+1)\tau_i^j + 4 {\cal{H}}a\eta^{kj}\partial_{(i}S_{k)}=0.
  \end{eqnarray}
\\
\\
\emph{Tensor}
  \begin{equation}\label{Vis tensor pert}
    \chi_i^{j\prime\prime} + 2{\cal{H}}\chi_i^{j\prime} - \nabla^2 \chi_i^j + \frac{1}{2} m^2 a^4(a^2+1)\chi_i^j  = 0.
  \end{equation}

Similarly to the tensorial equations for the GRT and for the
scalar-tensor theory, this last equation describe the evolution of
free GWs related to the NP mode $\Psi_4$. But now it is the new
term which contains $m^2$ that contributes to the parametric
amplification of GWs.

The equations for the scalar modes ($s = 0$) are now given by the
set (\ref{Vis scalar 1}), (\ref{Vis scalar 2}) and (\ref{Vis
scalar 3}). The argumentation for the scalar modes $\pi_{ij}$ and
$\psi_{ij}$ is similar to the case of the scalar-tensor theories,
except for the absence of the scalar field perturbation $\delta
\varphi$. Again, the three equations must be solved simultaneously
in order to find the evolution of the GWs with helicity $s = 0$
and to find the evolution of the density perturbation which is
related to the metric perturbations through the equation (\ref{Vis
scalar 1}):
\begin{equation}
\frac{\delta \rho}{\rho} = \frac{\delta \rho}{\rho}(\phi,B,\xi^i_i).
\end{equation}

Conversely, we have now GWs with helicity $s = \pm 1$ which
correspond to the mode $\Psi_3$. The equation which describes the
evolution of the GW amplitudes for this mode is the equation
(\ref{Vis vector 2}). Note the similarity of this equation to the
equation (\ref{Vis tensor pert}), except for the presence of the
term containing $S^i$ in the equation (\ref{Vis vector 2}). The
presence of this term makes the vector GW modes coupled with the
vector perturbations since $S^i$ is coupled with the fluid vector
perturbations through equation (\ref{Vis vector 1}). Furthermore,
the perpendicular part of the vector perturbation (which is a pure
vector) is a function of the quantities $S^i$ and $\tau^{ij}$:
\begin{equation}
V^i_{\perp} = V^i_{\perp}(S^i,\tau^{ij}).
\end{equation}

Regarding the CMB anisotropy, the presence of the longitudinal
vector modes of GWs do not yield the well know version of the
Sachs-Wolfe effect which appears in the GRT or in the
scalar-tensor theory \cite{Giovannini2005}. Theories which present
vector GWs ($\Psi_3 \neq 0$) give rise to a nontrivial Sachs-Wolfe
effect which leaves a vector signature of the quadrupolar form
$Y_{2,\pm 1}$ on the CMB polarization (see detailed discussion in
\cite{Bessada2008}).

\section{Final Remarks}\label{sec 6}

In the present work we have studied the evolution equations of
cosmological GWs in alternative theories of gravity. Since the
most part of the alternative theories present more than the two
usual $+$ and $\times$ polarizations of GWs, we have addressed the
problem of how one could take into account the new polarization
states in the cosmological metric perturbations.

First of all, we have presented an overview of the NP formalism
since it is particularly suitable for evaluating the number of
non-null GW modes of any theory. Then, we have proposed that the
construction of the metric perturbations for a given theory should
depend on the number of non-null NP parameters of the theory.

The formalism developed here is quite general and can be applied
for a wide range of alternative theories of gravity. In order to
show that, we have evaluated the evolution equations for GWs for
two different theories: a class of scalar-tensor theories, for
which Brans-Dicke theory is a particular case, and a massive
bimetric theory. In the first case, the theory presents two scalar
modes ($s = 0$) and two tensor modes ($ s = \pm 2$) of GWs. In the
case of the bimetric theory, GWs have in addition the two vector
modes ($s \pm 1$), totalizing six polarization states of GWs,
i.e., the most general case in the context of a four-dimensional
theory of gravity.

A qualitative analysis of the governing equations of the scalar
perturbations have shown that the evolution of the density
perturbations of the cosmological fluid depends not only on the
generalized Newtonian potential but also on the amplitudes of the
scalar GWs. Another direct consequence of the presence of
$\pi_{ij}$ and $\psi_{ij}$ is the possible signatures that this
quantities would leave in the angular pattern of the CMB. Such a
signature might impose strong limits in the amplitudes of the
scalar GWs by the analysis of the CMB data. Moreover, a remarkable
effect on the CMB which have already been studied in the
literature \cite{Bessada2008} is a non-usual Sachs-Wolfe effect
which appears due the presence of the longitudinal vector GW modes
on the geodesic equation for photons. Such vector GW modes are
present, for example, in the massive bimetric theory analyzed
here.

The detection of GWs is a particularly challenging issue and it
may be the final answer to the ``correct'' theory of gravity. If
GWs present non-tensorial polarization modes as discussed in the
present paper, we will have a stochastic cosmological background
of GWs which is a mixture of all the polarization modes. If, in
analyzing such a background, scalar and/or vector GWs could be
found, the result would be disastrous for the Einstein theory.

The evaluation of the response function of the non-tensor
polarization modes for interferometric GW detectors was carried
out in the references \cite{Nishi2009} and \cite{Corda2009}.
Particularly, Nishizawa et al. \cite{Nishi2009} have found that
more than three detectors can separate the mixture of polarization
modes in the detector outputs. But they have considered only
separation between the three groups: scalar, vector and tensor
modes of GWs. Furthermore, they have found that, statistically,
the GW detectors have almost the same sensitivity to each
polarization mode of the stochastic background of GWs. In the work
by Corda \cite{Corda2007}, the dectability of a particular
polarization was discussed, namely, the longitudinal scalar
component. It was also shown that the angular dependence of such a
mode could, in principle, allows discriminating this polarization
with respect to that of GRT.

A positive detection for certain modes and a negative detection
for others may exclude a particular theory or, at least, establish
strong constraints to the alternative theories of gravity. But it
is important to emphasize that, the confirmation by the
observation of the number of non-null GW modes is not enough to
determine the ``correct'' theory, since a number of theories can
have the same number of non-null modes. It is also necessary to
evaluate the spectrum of GWs for each mode and for each theory.
The evaluation of the spectrum can only be done following the
cosmological evolution of the GWs. In this sense, the equations
given in this paper are the first step for such a computation.
Finally, the analysis of the response of a given mode and the
evaluation of the spectrum can bring crucial answers for the
comprehension of the gravity in cosmological scales.

\section*{Acknowledgments}
The authors would like to thank D. Bessada for helpful discussions
and the referees whose comments and criticisms help to improve
significantly the first version of the paper. MESA would like to
thank also the Brazilian Agency FAPESP for support (grant
06/03158-0). ODM and JCNA would like to thank the Brazilian agency
CNPq for partial support (grants 305456/2006-7 and 307424/2007-3
respectively).




\section*{References}
\begin{thebibliography}{10}


\bibitem{Eardley1973} Eardley, D.M., Lee, D.L., and Lightman, A.P., 1973, \emph{Phys. Rev. D} 8, 3308.
\bibitem{Newman} Newman, E., and Penrose, R., 1962, \emph{J. Math Phys.} 3,
566; see errata, \emph{ibid.} 1962, 4, 998
\bibitem{Lifshiftz1946} Lifshiftz, E.M., 1946, \emph{Zh. Eksp. Teor.
Phys.} 16, 587
\bibitem{Bardeen1980} Bardeen, J.M., 1980, \emph{Phys. Rev. D} 22, 1882
\bibitem{Peebles1993} Peebles, P.J.E., 1993, Principles of
Physical Cosmology (Princeton Univ. Press, Princeton)
\bibitem{Mukhanov1992} Mukhanov, V.F., Feldman, H.A., and
Brandenberger, R.H., 1992, \emph{Phys. Reports} 215, 205
\bibitem{Durrer2004} Durrer, R., 2004, Cosmological
Perturbation Theory, Lecture Notes in Physics, Edited by E.
Papantonopoulos (Springer, Berlin)
\bibitem{Malik2009} Malik, K.A., and Wands, D., 2009, \emph{Phys.
Reports} 475, 1
\bibitem{Grishchuk1974} Grishchuk, L.P., 1974 \emph{Zh. Eksp. Teor.
Fiz.} 67, 825
\bibitem{Barrow1993_2} Barrow, J.D., Mimoso, J.P., and de Garcia Maia,
M.R., 1993, \emph{Phys. Rev. D} 48, 3630
\bibitem{Maia1994} de Garcia Maia, M.R., and Barrow, J.D., 1994, \emph{Phys. Rev.
D} 50, 6262
\bibitem{Capozziello2007} Capozziello, S., Corda, C., and de
Laurentis, 2007, \emph{Mod. Phys. Lett. A} 22, 2647
\bibitem{Alves2009B} Alves, M.E.S., Miranda, O.D., and de Araujo,
J.C.N., 2009, \emph{Phys. Lett. B} 679, 401
\bibitem{Capozziello2008} Capozziello, S., Corda, C., and De
Laurentis, M.F., 2008, \emph{Phys. Lett. B} 669, 255
\bibitem{Corda2008} Corda, C., 2008, \emph{Astropart. Phys.} 30,
209
\bibitem{Bessada2008} Bessada, D., and Miranda, O.D., 2008, \CQG 26, 045005
\bibitem{Brans1961} Brans, C., and Dicke, R.H., 1961, \emph{Phys. Rev.}
124, 925
\bibitem{Casas1991} Casas, J.A., Garcia-Bellido, J., and Quirós,
M., 1991, \emph{Nucl. Phys. B} 361, 713
\bibitem{vis1998} Visser, M., 1998, \emph{Gen. Relativ. Gravit.} 30, 1717
\bibitem{Alves2006} Alves, M.E.S., Miranda, O. D., and de Araujo, J.C.N, 2007, \emph{Gen. Relativ. Gravit.} 39, 777
\bibitem{Alves2009} Alves, M.E.S., Miranda, O. D., and de Araujo, J.C.N,
2009, arXiv:0907.5190v1
\bibitem{Weinberg1972} Weinberg, S., 1972, Gravitation and
cosmology: principles and applications of the general theory of
relativity, New York: John Wiley $\&$ Sons
\bibitem{Will2006} C. M. Will, Living Reviews in Relativity (2006)
http://relativity.livingreviews.org/Articles/Irr-2006-3
\bibitem{Wagoner1970} Wagoner, R.V., 1970, \emph{Phys. Rev. D} 1,
3209
\bibitem{Nordtvedt1970} Nordtvedt, K., 1970, \emph{Astrophys. J.}
161, 1970
\bibitem{Maggiore2000} Maggiore, M., Nicolis, A., 2000, \emph{Phys. Rev.
D} 62, 024004
\bibitem{Rosen1973} Rosen, N., 1973, \emph{Ann. of Phys.} 84, 455
\bibitem{Will1993} Will, C.M., 1993, Theory and experiment in
gravitational physics, Cambridge: Cambridge University Press
\bibitem{vDVZ1970} van Dam, H., Veltman, M., 1970, \emph{Nucl. Phys.
B} 22, 397
\bibitem{Zakharov1970} Zakharov, V.I., 1970, \emph{JETP Lett.} 12, 312
\bibitem{paula2004} de Paula, W.L.S., Miranda, O.D. and Marinho, R.M., 2004, \CQG 21, 4595
\bibitem{Bicak2005} Bicak, J., and Katz, J., 2005, \emph{Czech. J. Phys.} 55, 105
\bibitem{Bicak2007} Bicak, J., Katz, J., and Lynden-Bell, D., 2007,
\emph{Phys. Rev. D} 76, 063501
\bibitem{Giovannini2005} Giovannini, M., 2005, \emph{IJMP D} 14,
363
\bibitem{Nishi2009} Nishizawa, A., Taruya, A., Hayama, K.,
Kawamura, S., and Sakagami, M., 2009, \emph{Phys. Rev. D} 79,
082002
\bibitem{Corda2009} Corda, C., 2009, arXiv:0905.2502
\bibitem{Corda2007} Corda, C., 2007, \emph{Astropart. Phys.}
28, 247












\end{thebibliography}
\end{document}